# ORGANIZATION AND DISSIPATION OF CURRENTS IN NON-EQUILIBRIUM THERMODYNAMICS

E. G. D. Cohen

*The Rockefeller University, 1230 York Avenue, New York, NY 10065 USA,*

*Department of Physics and Astronomy, The University of Iowa, Iowa City, IA, 52242*

---

## ABSTRACT

A dynamical theory for the organization and dissipation of a current in a non-equilibrium fluid near equilibrium is presented. This is based on the Lyapunov exponents of the phase space of the system.

---

This note is a sequel to a previous paper [1] on the organization of a current in a non-equilibrium fluid near equilibrium, e.g., in local equilibrium. In that paper it was argued that the most important aspect of a fluid in a non-equilibrium state is the *organization* of the flowing current, which is necessary to obtain the maximum entropy of the system when the system returns to the equilibrium state.

A current necessarily has an organized structure, since the fluid particles flow, on average, in a certain direction. This is completely opposite to what happens in thermal equilibrium, where the particle's velocities are Maxwellian, i.e., Gaussian or randomly distributed (consistent with the principle of maximum entropy of the equilibrium state).

It was also pointed out in [1], that the textbooks on non-equilibrium thermodynamics (e.g., [2]) exclusively consider the entropy production, i.e., the dissipation of the flowing currents, and do not consider the much more important self-organization of the current, mathematically represented as a vector [2]. In this note, some further consequences of the current's organization will be discussed.

External forces must be applied to bring a system in an equilibrium state into a non-equilibrium state. These forces can induce a gradient of e.g. the system's temperature, leading to a heat current, i.e. bringing it into a non-equilibrium state. The ensuing entropy production then causes a contraction of the phase space volume of the flowing fluid, which is a consequence of the necessary dissipation, as required by the second law of thermodynamics. It is well-known that a dynamical origin can be given for this phase space contraction using the (non-equilibrium) Lyapunov exponents of a non-equilibrium fluid [3].

In the case of "thermal equilibrium", denoted by a superscript "e", the sum of the Lyapunov exponents, $\lambda_i^e, \, i = 1 \ldots 6N,$ of the phase space of a fluid of $N$ particles vanishes, since they occur in positive and negative pairs, denoted, respectively by $\lambda_{i+}^e$ and $\lambda_{i-}^e$, and have equal magnitudes. Therefore, the phase space volume remains constant in equilibrium, i.e.,

$$\sum_{i=1}^{6N} \lambda_i^e = \sum_{i=1}^{3N} \left( \lambda_{i+}^e + \lambda_{i-}^e \right) = 0. \tag{1}$$

To the contrary, in non-equilibrium, denoted by the superscript "ne", the sum of the pairs of the Lyapunov exponents is negative, i.e.,

$$\sum_{i=1}^{6N} \lambda_i^{ne} = \sum_{i=1}^{3N} \left( \lambda_{i+}^{ne} + \lambda_{i-}^{ne} \right) < 0, \tag{2}$$

since $\lambda_{i-}^{ne} < 0$, and $\left| \lambda_{i-}^{ne} \right| > \lambda_{i+}^{ne}$, so that Eq. (2) implies a phase space contraction. The entropy production rate or dissipation rate, σ, is defined as

$$\sigma = \sum_{i=1}^{3N} \left| \lambda_{i+}^{ne} + \lambda_{i-}^{ne} \right|, \tag{3}$$

and must be positive [2]. The opposite signs of the $\lambda_{i+}^{ne}$ and $\lambda_{i-}^{ne}$ in Eq. (3) implies a disorganization of the current, due to its dissipation.

We will now give a corresponding discussion for the *organization* of the current in a non-equilibrium state. The application of the abovementioned external forces will induce a gradient [4], which in turn produces the current and changes the role of the equilibrium Lyapunov exponents in a fundamental way.

We propose that the organization (growth) rate of the current (ω) is given by the sum of the *differences* of corresponding positive and negative Lyapunov exponents, i.e., by

$$\omega = \sum_{i=1}^{3N} \left( \lambda_{i+}^{ne} - \lambda_{i-}^{ne} \right) = \sum_{i=1}^{3N} \left( \lambda_{i+}^{ne} + \left| \lambda_{i-}^{ne} \right| \right). \tag{4}$$

The same signs of the $\lambda_{i+}^{ne}$ and $\left| \lambda_{i-}^{ne} \right|$ on the right-hand side of Eq. (4), imply a cooperation of the Lyapunov exponents, leading to the current's organization. This shows that, since $\lambda_{i+}^{ne} \sim \left| \lambda_{i-}^{ne} \right|$ [5], the current organization rate, ω, is much larger than the current dissipation rate, σ, as it should be for a normal viscous fluid [6]. Therefore, Eqs. (3) and (4) imply a phase space *expansion*, which is much larger than the dissipative phase space contraction.

Summarizing the new role of the *physical* organization of the current in a non-equilibrium state we have:

1.  It provides the source of the equilibrium entropy [1], when a non-equilibrium system goes to equilibrium; and

2.  It causes an *expansion* of the fluid's phase space due to the current's organization rate which overwhelms the phase space contraction rate of the fluid due to the current's dissipation, as it should be.

## <u>References and Endnotes</u>

[1]     E. G. D. Cohen and R. L. Merlino, Mod. Phys. Lett. B **28**(9), 1450073 (2014).

[2]     S. R. de Groot and P. Mazur, *Nonequilibrium Thermodynamics* (North Holland, Amsterdam, 1962).

[3]     D. J. Evans and G. P. Morriss, Statistical Mechanics of Nonequilibrium Liquids (ANU E Press, Canberra, 2007).

[4]     E.g., the application of a pressure difference to a fluid in a pipe produces a Poiseuille flow.  L. D. Landau and E. M. Lifshitz, vol.6 *Fluid Mechanics* p.55 (Pergamon Press, 1959).

[5]     D. J. Evans, E. G. D. Cohen, and G. P. Morriss, Phys. Rev. A **42**, 5990 (1990).

[6]     As shown in [5], the sum of the differences between the positive and negative Lyapunov exponents, which represent the current organization rate, are an order of magnitude larger than the entropy production.